\PassOptionsToPackage{dvipsnames}{xcolor}
\documentclass[sigconf]{acmart}

\usepackage{booktabs} 

\usepackage[english]{babel}
\usepackage[utf8]{inputenc}
\usepackage{pifont}
\usepackage{xspace}
\usepackage{graphicx}
\usepackage{textcomp}
\usepackage{listings}
\usepackage{todonotes}
\usepackage{booktabs}
\usepackage{makecell}
\usepackage[framemethod=TikZ]{mdframed}
\setcopyright{none}

\definecolor{verylightgray}{rgb}{.97,.97,.97}
\newcommand{\cmark}{\textcolor{ForestGreen}{\ding{51}}\xspace}
\newcommand{\xmark}{\textcolor{WildStrawberry}{\ding{55}}\xspace}
\newcommand{\btwo}{D1} 
\newcommand{\paper}{D2}
\newcommand{\bthree}{D3}

\mdfdefinestyle{answerbox}{%
          skipabove=\topskip,
          roundcorner=5pt,%
          shadow=false,%
          linecolor=CadetBlue,%
          linewidth=1pt,%
          backgroundcolor=verylightgray,%
}

\acmConference[SAC '24]{The 39th ACM/SIGAPP Symposium on Applied Computing}{April 8--12, 2024}{Avila, Spain}
\acmDOI{10.1145/3605098.3636060}
\acmISBN{979-8-4007-0243-3/24/04}

\acmArticle{4}
\acmPrice{15.00}

\begin{document}
\title{Explainable Ponzi Schemes Detection on Ethereum}
  
\author{Letterio Galletta}
\orcid{0000-0003-0351-9169}
\affiliation{%
  \institution{IMT School for Advanced Studies Lucca}
  \city{Lucca} 
  \country{Italy}
}
\email{letterio.galletta@imtlucca.it}

\author{Fabio Pinelli}
\orcid{0000-0003-1058-6917}
\affiliation{%
  \institution{IMT School for Advanced Studies Lucca}
  \city{Lucca} 
  \country{Italy}
}
\email{fabio.pinelli@imtlucca.it}





\begin{abstract}

Blockchain technology has been successfully exploited for deploying new economic applications. 
However, it has started arousing the interest of malicious actors who deliver scams to deceive honest users and to gain economic advantages. Ponzi schemes are one of the most common scams.
Here, we present a classifier for detecting smart Ponzi contracts on Ethereum, which can be used as the backbone for developing detection tools.
First, we release a labelled data set with 4422 unique real-world smart contracts to address the problem of the unavailability of labelled data.
Then, we show that our classifier outperforms the ones proposed in the literature when considering the AUC as a metric.
Finally, we identify a small and effective set of features that ensures a good classification quality and investigate their impacts on the classification using eXplainable AI techniques.

\end{abstract}



\maketitle

\section{Introduction}\label{sec:intro}

Blockchain is revolutionizing how individuals and companies exchange digital assets without the control of a central authority.
This technology has been successfully exploited for deploying new economic applications, e.g., cryptocurrencies~\cite{Nakamoto08} and DeFi~\cite{defiforbes}.
However, soon after this technology became widespread and its economic value increased, it has started
arousing the interest of malicious actors who are eager to take advantage due to the pseudonymity of these platforms and the lack of regulation~\cite{Moore13}:
on the one hand, they exploit cryptocurrencies to transfer currency without being tracked by authorities;
on the other hand, they deliver scams to deceive honest users willing to make revenues through cryptocurrencies.
Nowadays, many types of scams can be found on blockchain platforms, such as exploits, hacks, and phishing~\cite{Bartoletti2021Access}:
estimates say that scams in Bitcoin~\cite{Vasek15} gathered more than 7 million USD.

Among the various scams, Ponzi schemes have approached the blockchain world, first on Bitcoin~\cite{Vasek15} and more recently on Ethereum~\cite{BartolettiCCS20}.
These are fraudulent investment operations where older investors obtain returns from new investors' money rather than legitimate business activities.
Although the actual conditions to gain money depend on the specific rules of the scheme, a common feature is that participants who want to redeem their investments have to make new participants join the scheme.
Participants who join later are the most likely to lose their money.
Thus, the development of automatic techniques to counter these scams is required to protect average users and to allow them to participate safely in the blockchain economy.

This paper focuses on Ethereum and smart contracts to deliver Ponzi schemes, called smart Ponzi contracts.
We implement an automatic technique for classifying smart contracts, which can be used as the backbone for developing new detection tools.
More precisely, we provide the following contributions.
First, we address the problem of the unavailability of public data sets to train effective automatic classifiers. 
Thus, we release a reusable data set that collects 4422 unique real-world smart contracts, where 3749 (84.78\%) are not-Ponzi, and 673 (15.22\%) are Ponzi. 
Our data set contains both information about the transaction history of the contracts as well as their bytecode.
Then, we implement a binary classification model to detect smart Ponzi contracts.  
Our experiments show that the proposed model performs better than the models proposed in the literature when considering the AUC as a metric and achieves high accuracy for practical use.
Finally, we address the issue of identifying a small and effective set of features that ensures a good quality of the classification process.
To find this set, we proceed as follows: first, we introduce new features and show through experiments that they improve the classification.
Then, we consider the union of our features and those from the literature and identify which ones can be safely removed because they do not contribute to classification.
We adopt eXplainable AI (XAI) techniques to investigate the contribution of each feature, which, to the best of our knowledge, has never been done so far in this context. 

In summary, the main contributions of this paper are:
\begin{itemize}
\item a reusable and publicly available data set that collects 4422 real-world smart contracts where 3749 are not Ponzi, and 673 are Ponzi;
\item a binary classifier to detect smart Ponzi contracts that outperform  classifiers in the literature when considering the AUC as a metric;
\item a small and effective set of features that ensures a good classification quality; 
\item the study of the impact of such features on the classification using eXplainable AI techniques.
\end{itemize}

\paragraph*{Structure of the paper}
We proceed as follows. 
First, we introduce Ponzi schemes and smart Ponzi contracts (Section~\ref{sec:back}) and discuss the connection between our work and the literature (Section~\ref{sec:related}). 
Then, we describe our data set and the features we use (Section~\ref{sec:dataset}).
In Section~\ref{sec:experiments}, we build our binary classifier and perform an experimental evaluation to study its quality and the impact and importance of the features. 
Finally, we draw some conclusions and discuss future work in Section~\ref{sec:concl}.

\paragraph*{Availability} 
The data set and the source code used for the experiments presented in this paper are available online.\footnote{\url{https://github.com/fpinell/ponzi_ml}}

\section{Background: Smart Ponzi contracts}\label{sec:back}

Ponzi schemes are classic frauds concealed as ‘‘high-yield’’
investment programs. 
The initiator of the scheme generates returns for existing investors through revenue paid by new investors rather than from legitimate business 
activities or profits of financial trading. 
More in general, the U.S. Securities and Exchange
Commission~\cite{secPonzi}
defines Ponzi schemes as
``an investment fraud that involves the payment
of purported returns to existing investors from funds contributed
by new investors. Ponzi scheme organizers often solicit new investors by promising to invest funds in opportunities that generate high returns with little or no risk. With little or no
legitimate earnings, Ponzi schemes require a constant flow of
money from new investors to continue. Ponzi schemes inevitably
collapse, most often when it becomes difficult to recruit new
investors or when a large number of investors ask for their funds
to be returned.''

Although the actual conditions to gain money depend on the specific rules of the scheme, a user who wants to redeem her investment has to make new users join the scheme. 
In this way, the schemes create a pyramid of investors, where the initiator is at the top, and the investors at level $l+1$ compensate for the investment of those at level $l$.  
Once a scheme collapses, investors at the top of the pyramid gain money, while those at the bottom lose it.

The spread of cryptocurrency and smart contracts has created new opportunities to 
deploy this kind of fraud.
Indeed, it is possible to find samples of smart contracts implementing Ponzi schemes, called smart Ponzi contracts, deployed on the main blockchain platforms like Ethereum and Bitcoin.
This paper focuses on the Ethereum platform.
According to the literature~\cite{Bartoletti2018DataMF} smart Ponzi contracts have several attractive features to be used for scams:
\begin{enumerate}
\item The initiator of a smart Ponzi could stay anonymous,
since deploying the contract on the blockchain and withdrawing money from
it only requires an Ethereum account that does not reveal her real identity.
\item Once deployed on the blockchain, smart contracts are ‘‘unmodifiable’’ and ‘‘unstoppable’’. 
Thus, no central authority could terminate the execution of the scheme, seize the money, and refund the victims.
\item Since the code of smart contracts is public, immutable, and its execution is automatically enforced by the blockchain platform, investors may believe that no one can take advantage of their money and that they could eventually gain the declared interests.
\end{enumerate}
The most significant feature of a smart Ponzi is the policy used to redistribute new investments among participants, i.e., how the money flows. This requires a smart Ponzi to maintain a data structure storing participants’ information and implement a strategy for redistributing dividends. 

Identifying redistribution behaviour is crucial to classify a contract as a smart Ponzi.
Also, it is challenging because many other kinds of contracts, e.g.,  gambling games, may have similar behaviour, which may induce many false positives.
Bartoletti et al.~\cite{BartolettiCCS20} proposed the following four requirements to classify a smart contract as a Ponzi scheme:
\begin{description}
\item[R1] the contract redistributes money to the investors according to a given logic;
\item[R2] the contract receives money only from the investors;
\item[R3] each investor makes a profit if a certain number of investors join subsequently the contract investing some money;
\item[R4] the later an investor joins the contract, the higher the risk of losing her money.
\end{description}
A smart contract is classified as a smart Ponzi when it satisfies \emph{all} four requirements.
Note that requirement R1 rules out contracts that provide users with some assets but do not implement a distribution logic to participants, e.g., tokens; 
requirement R2 ensures that a participant invests a certain amount in joining the contract;
requirement R3 demands a constant flow of new investments for investors to make a profit;
requirement R4 characterizes the fraudulent nature of smart Ponzi contracts because it reflects the fact that making a profit for investors is likely impossible after a certain point in time: 
too many victims must join the scheme for the contract to have enough money to reward all the participants.
Thus, the scheme collapses when this happens.
Note that the requirements above impose no condition on the money received or not by the initiator of the scheme. 
We will study the need for such a condition in our experimental evaluation.

\section{Related work}\label{sec:related}
Since the inception of Bitcoin, cryptocurrencies and blockchain systems have attracted the attention of cybercriminals, who exploit them to carry out potentially untraceable scams.
Since the entire transaction history is publicly available and provides accurate records of user behaviours, several papers~\cite{Bartoletti2021Access,Liu2021Access,trozze2022cryptocurrencies} have proposed machine learning techniques to detect possible frauds and scams. 
Below, we discuss the proposals that are the most similar to ours.

Bartoletti et al.~\cite{Bartoletti2018DataMF} study the problem of identifying 
Ponzi schemes in Bitcoin through data mining.
They released a public data set of Bitcoin addresses and an open-source tool to build such a data set;
then, they apply different classification algorithms and systematically evaluate them to identify the best discriminating features for detecting Ponzi schemes in Bitcoin.
Our work shares their goal of providing a public data set of smart Ponzi contracts and a good classifier and of identifying the most discriminating features. 
However, we target Ethereum smart contracts and consider different classifiers.
Moreover, we use XAI techniques to study the contributions of the features.

Bartoletti et al.~\cite{BartolettiCCS20} consider smart Ponzi contracts in Ethereum. 
First, they define four criteria based on behavioural aspects to identify a contract as a smart Ponzi and produce a data set with several contracts satisfying such criteria. 
Then, they perform several manual analyses on some contracts, including the security of their code and the fairness of the distribution policy. 
In our work, we adopted their criteria, enriched their data set by considering new features, and added new contracts. 
Moreover, we trained and experimented with different classifiers to automatically determine if a smart contract is a Ponzi scheme. 

Chen et al.~\cite{Chen2019} provide a reusable data set with real-world samples and evaluate different classifiers. 
They use two classes of features: the account features taken from the transaction history and code features extracted from the contract's bytecode. 
We follow the same approach for building and evaluating different classifiers. 
Moreover, we reused and extended their data set with new features and contracts. 
In contrast, we provide a more accurate analysis and an explanation of how the different features impact the classification process, and we show that our best classifier outperforms theirs. 

Chen et al.~\cite{Chen2021} propose SADPonzi, a detection tool based on symbolic execution. 
The tool analyzes the bytecode of contracts to extract semantic information and to identify investor-related transfer behaviours and the distribution strategies adopted by the scheme. 
SADPonzi performs the classification only by looking at the code, not the transactions' history. 
We include in our dataset some contracts identified by this tool.
The main differences with our work are that we resort to machine learning techniques to classify contracts and do not consider bytecode but only the information about the transaction history.

Wang et al.~\cite{WANG2021} propose an approach based on Long-short Term Memory Network to detect smart Ponzi.
In contrast, our classifier is not based on neural networks, and it is obtained after an experimental evaluation that tests different models and hyper-parameters. 
Moreover, we provide a precise account of how the various features impact the classification, and our data set is publicly available. 

Fan et al.~\cite{fan2020} propose PonziTect, a detection method based on ordered boosting that classifies contracts considering only their bytecode.
They adopt data augmentation to solve the problem of imbalanced data by increasing the proportion of smart Ponzi at the boundary.
Here, we do not consider the problem of imbalanced data because the algorithms we use are not affected too much by this issue. 
Another difference is that we focus only on transaction history instead of bytecode and that we release the data set and code we used to train and test our models.

Lou et al.~\cite{lou2020} use convolutional neural networks and focus mainly on bytecode features. The pipeline they propose is standard:
first, they transform smart contracts into single-channel images and then adopt the spatial pyramid pooling method to ensure that the generated images have the same size. 
In contrast, our classifier is not based on neural networks and considers only account features.
Moreover, here we adopt XAI techniques to understand how the various features impact the classification process.
In addition, our data set is public. 
 
Ibba et al.~\cite{ibba2021} build a machine learning model that uses transaction features,  the bytecode, and the Solidity source code.
They tested their approach with decision trees, support vector machines, and naive Bayes.
In contrast, our classifier considers only transaction features, and we study how the various features impact the classification process.
Moreover, our data set is publicly available. 

As a last observation, none of the papers above uses XAI techniques to study the contribution of the features. 

\section{Data set construction}\label{sec:dataset}

This section describes the data and features we use to detect smart Ponzi.
We build two data sets from the same sources but based on different features. 
Below, after describing our data and features, we present a qualitative analysis of some features to understand how they distribute across the two classes and how much they may discriminate during the classification.  

 \subsection{Feature description}
We built our data set based on others from the literature~\cite{BartolettiCCS20,Chen2019,Chen2021}.
Specifically, we adopted their labelling results, which involved the manual inspection of each contract code in verifying their compliance with the requirements R1-R4 outlined in Section~\ref{sec:back}.
The resulting data set contains 4422 smart contracts, with 3749 (85.23\%) labelled as not-Ponzi and 673 (14.77\%) as Ponzi. 
Nonetheless, we have introduced additional features designed to capture novel characteristics, which, in turn, enhance the performance of our classifier. For more details, please refer to Section~\ref{sec:experiments}.

Below, we report the list of the features for each contract: 
\begin{enumerate}
    \item \emph{Address}: the address of the smart contract (not used for classification);
    \item \emph{Balance}: the amount of currency in Ether (ETH) deposited in the contract;
    \item \emph{Lifetime}: the difference between the time of the first and the last transaction made or received;
    \item \emph{Tx\_{in}}: the number of input transactions;
    \item \emph{Tx\_{out}}: the number of output transactions;
    \item \emph{Investment\_{in}}: the number of input transactions that deposit an amount of ETH in the contract;
    \item \emph{Payment\_{out}}: the number of output transactions paying an amount of ETH; 
    \item \emph{\#addresses\_{paying}\_{contract}}: the number of distinct addresses (an address identifies a user) that paid the contract;
    \item \emph{\#addresses\_paid\_by\_contract}: the number of distinct addresses paid by the contract;
    \item \emph{Mean\_v1}: the average of the differences of the number of input/output transactions from/to the same address;  
    \item \emph{Mean\_v2}: the average of the differences of the amount of ETH received and paid by the contract involving the same address;
    \item \emph{Sdev\_v1}: the standard deviation of the differences in the number of input and output transactions involving the same address;
    \item \emph{Sdev\_v2}: the standard deviation of the differences between the amount of ETH in and out involving the same address;
    \item \emph{Paid\_{rate}}: the ratio between Tx\_{in} and Tx\_{out};
    \item \emph{Paid\_{one}}: the ratio between the number of investors paid many times and the number of total investors;%
    \item \emph{Know\_rate}: the proportion of receivers who have invested before payment;
    \item \emph{N\_maxpayment}: the max number of payments to all participants;
    \item \emph{Skew\_v1}: the skewness of the differences of the number of input and output transactions involving the same address;
    \item \emph{Skew\_v2}: the skewness of the differences of the amount of ETH in and out involving the same address;

    \item \emph{Investment\_{in}/Tx\_{in}}: the ratio between Investment\_{in} and Tx\_{in};
    \item \emph{Payment\_{out}/Tx\_{out}}: the ratio between Payment\_{out} and Tx\_{out};   
    \item \emph{Percentage\_{some}\_{tx}\_{in}}: the percentage of active days with at least one input transaction during the contract lifetime;
    \item \emph{Sdev\_{tx}\_{in}}: the standard deviation of the number of transactions per day;
    \item \emph{Percentage\_{some}\_{tx}\_{out}}: 
    the percentage of active days with at least one output transaction during the contract lifetime;
    \item \emph{Sdev\_{tx}\_{out}}: the standard deviation of the number of transactions in output per day;
    \item \emph{Initiator\_{get}\_{eth}\_{wo}\_{investing}}: this feature is 1 if the contract initiator has earned ETH without any investment, 0 otherwise;  
    \item \emph{Initiator\_{get}\_{eth}\_{investing}}: this feature is 1 if the initiator has earned ETH investing in the contract, 0 otherwise;
    \item \emph{Initiator\_{no}\_{eth}}: this feature is 1 if the Initiator has obtained no ETH investing in the contract, 0 otherwise.
\end{enumerate}

The first 19 features are inherited from previous works, while the last nine are the new ones we introduced.
For example, Feature 15 estimates how often a contract interacts with users it already knows. 
A high value of this feature means more interactions (we expect it to happen often for smart Ponzi). We now briefly comment on the new features.

Features 20 and 21 aim to capture requirement R1 of Section~\ref{sec:back} by measuring the percentage of transactions distributing Ether among investors.
Since we expect that not-Ponzi contracts present a lower percentage than Ponzi ones, these features are uniquely based on currency exchanges.
Features 22 and 24 monitor the number of input and output transactions per day over time.  
A small value of Feature 22 (respectively 24) indicates that the contract was active for a few days considering input transactions (output, respectively, for Feature 24). 
On the contrary, a high value means the contract presented a more regular activity.
We also consider the standard deviation of daily transactions (Features 23 and 25) to capture the variability during the contract life. 
In particular, we expect that Ponzi contracts have a short lifetime in which they receive several user investments.
Indeed, since making a profit for investors is likely impossible after a certain time, the scheme collapses, and the contract will no longer receive new investments. 
These features try to capture this behaviour, also considering what is specified by requirements R2, R3, and R4 of Section~\ref{sec:back}.
The last three features, 26, 27, and 28, verify whether the initiator received money from the contract. 
Indeed, in smart Ponzi contracts, the initiator usually receives a certain amount of money, even without an initial investment. 
We add these features to study whether receiving a certain amount of money is peculiar to this kind of fraud.
We expect most contracts labelled Ponzi to have Feature 26 and Feature 27 equal to 1. 
On the contrary, Feature 28 equals 1 for non-Ponzi contracts since finding the initiator of a Ponzi that receives no Ether is quite unusual.
\begin{table}[tbp]
\centering\footnotesize
\caption{A summary of the features included in the various data sets. The new features are in \textbf{bold}.}
\label{tbl:dataset:features}
\begin{tabular}{lrrrrr}
{} &  \btwo &  \paper &  \bthree \\ \hline
Address                      &            \cmark &                  \cmark &            \cmark  \\
Balance                      &            \cmark &                  \cmark &            \cmark  \\
Lifetime                     &            \cmark &                   \xmark&            \cmark  \\
Tx\_in                        &            \cmark &                  \xmark &            \cmark \\
Tx\_out                       &            \cmark &                  \xmark &            \cmark \\
Investment\_in                &            \cmark &                  \cmark &            \cmark  \\
Payment\_out                  &            \cmark &                  \cmark &            \xmark  \\
\#addresses\_paying\_contract   &            \cmark &                  \xmark &            \cmark  \\
\#addresses\_paid\_by\_contract  &            \cmark &                  \xmark &            \cmark  \\
Mean\_v1                      &            \cmark &                  \cmark &            \cmark  \\
Mean\_v2                      &            \cmark &                  \cmark &            \cmark  \\
Sdev\_v1                      &            \cmark &                  \cmark &            \cmark  \\
Sdev\_v2                      &            \cmark &                  \cmark &            \cmark  \\
Paid\_rate                    &            \cmark &                  \cmark &            \cmark  \\
Paid\_one                     &            \cmark &                  \cmark &            \cmark  \\
Known\_rate                   &            \cmark &                  \cmark &            \cmark  \\
N\_maxpayment                 &            \cmark &                  \cmark &            \cmark  \\
Skew\_v1                      &            \cmark &                  \cmark &            \cmark  \\
Skew\_v2                      &            \cmark &                  \cmark &            \cmark  \\
\textbf{Investment\_in/tx\_in} &            \cmark &                  \xmark &            \cmark  \\
\textbf{Payment\_out/tx\_out} &            \cmark &                  \xmark &            \cmark  \\
\textbf{Percentage\_some\_tx\_in}        &            \cmark &                  \xmark &            \cmark  \\
\textbf{Sdev\_tx\_in}                   &            \cmark &                  \xmark &            \cmark  \\
\textbf{Percentage\_some\_tx\_out}       &            \cmark &                  \xmark &            \cmark  \\
\textbf{Sdev\_tx\_out}                  &            \cmark &                  \xmark &            \cmark  \\
\textbf{Initiator\_gets\_eth\_Wo\_investing}  &            \cmark &                  \xmark &            \cmark \\
\textbf{Initiator\_gets\_eth\_investing}     &            \cmark &                  \xmark &            \xmark  \\
\textbf{Initiator\_no\_eth}                 &            \cmark &                  \xmark &            \xmark  
\end{tabular}
\end{table}

We build two data sets \btwo\ and \paper\ to evaluate how the different features impact the classification quality. 
The dataset labelled as \btwo\ incorporates all the previously mentioned features, while the \paper\ dataset exclusively includes the features employed by Chen et al.~\cite{Chen2019}. We adopt their classifier as a baseline in our experiments to gauge the enhancements in classification performance.
Table~\ref{tbl:dataset:features} summarizes which features are present in the various data sets: symbol \cmark means that the feature is included, whereas \xmark means that the feature does not appear.
Note that the way we generate the data set \bthree\ will be discussed in Section~\ref{sec:experiments}, and that features \emph{lifetime}, \emph{tx\_in}, \emph{tx\_out}, \emph{\#addresses\_paying\_contract}, and \emph{\#addresses\_paid\_by\_contract} are not present in Chen et al.~but are in other papers.

\subsection{Qualitative analysis of the features}
\begin{figure*}[p]
    \centering
    \footnotesize
    \begin{tabular}{cccc}
    \includegraphics[width=0.85\columnwidth]{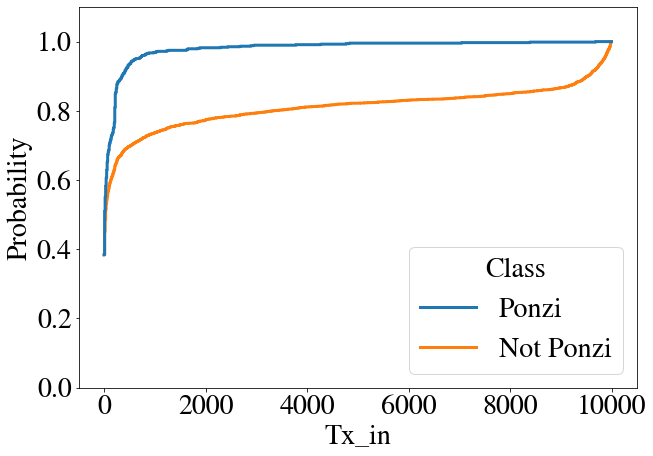} & 
    \includegraphics[width=0.85\columnwidth]{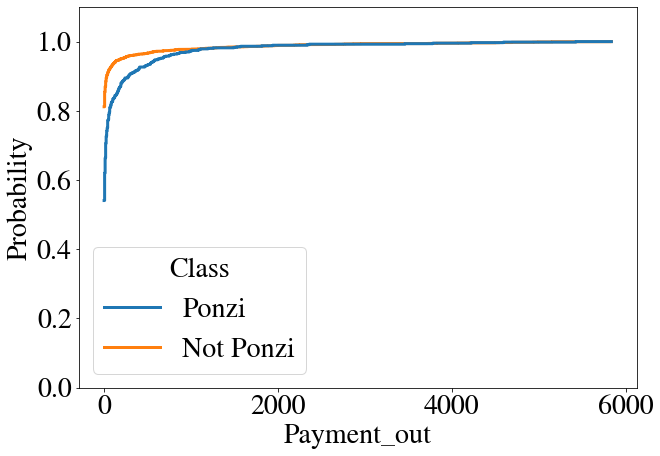} \\
    \includegraphics[width=0.85\columnwidth]{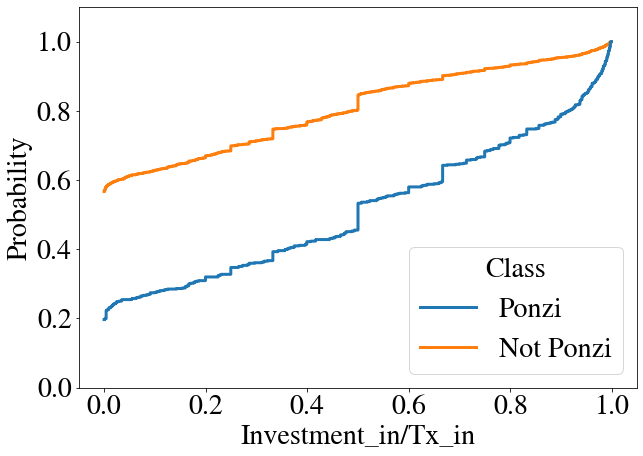} &
    \includegraphics[width=0.85\columnwidth]{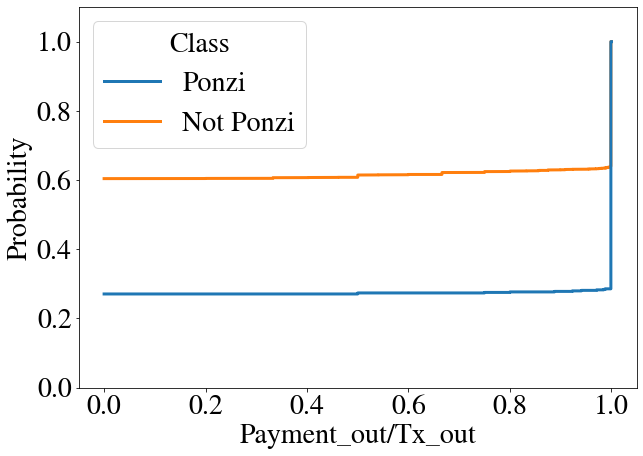} \\
    \includegraphics[width=0.85\columnwidth]{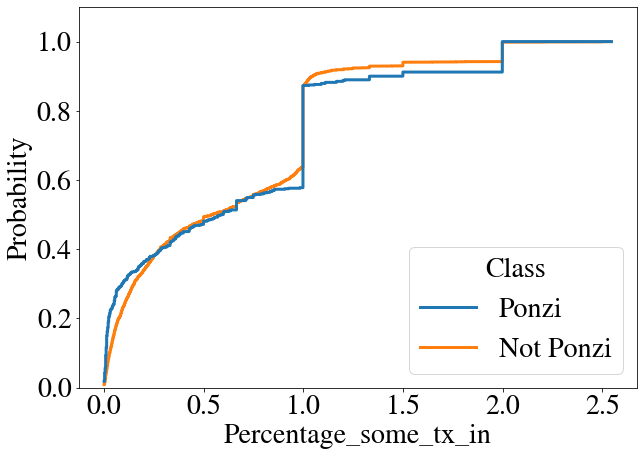} &
    \includegraphics[width=0.85\columnwidth]{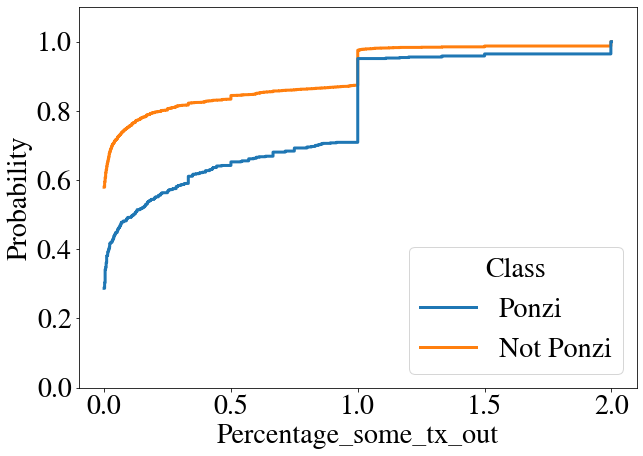} \\
    \includegraphics[width=0.85\columnwidth]{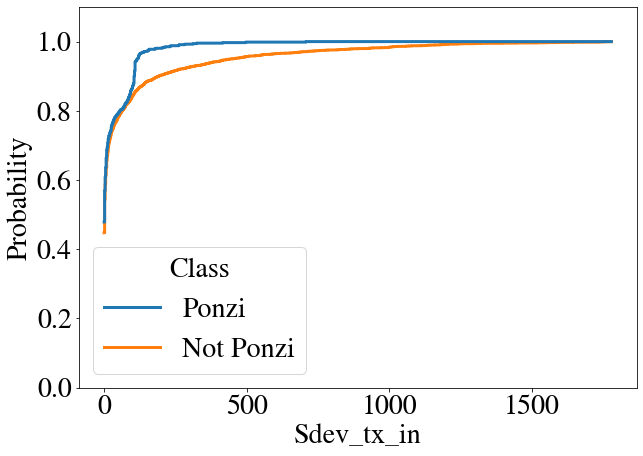} & 
    \includegraphics[width=0.85\columnwidth]{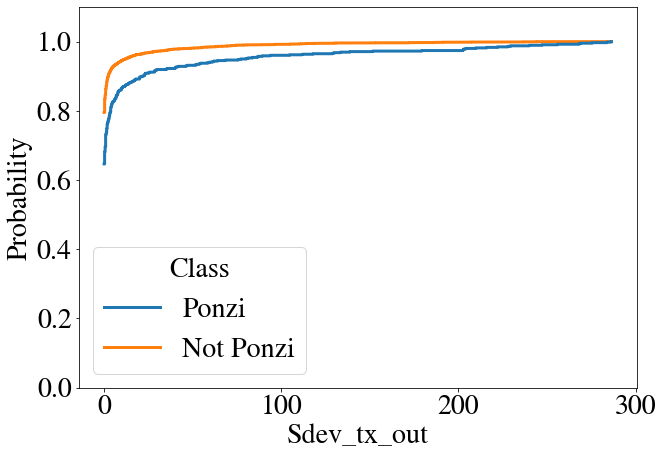} \\
    \end{tabular}
    \caption{The cumulative distributions of some continuous features: the distributions of the smart Ponzi are in blue, in orange for the not Ponzi ones.}
\label{fig:continuous_features}
\vspace{-3mm}
\end{figure*}

We report below a qualitative analysis of how the new features are distributed across the two classes and which seems to be the most characteristic for each class. 
We also consider Features 4 and 7 that will play a role during our experiments in Section~\ref{sec:experiments}.
Figure~\ref{fig:continuous_features} shows the cumulative distributions for the new continuous features (Features 20-25 above): the blue line represents the Ponzi behaviour, while the orange line represents the not-Ponzi. 
For visualization purposes, we discard the 1st and the 99th percentile in the plots; however, the shape of the cumulative distributions is preserved. 

The two populations present a different behaviour concerning the input transactions (\emph{Tx\_in}): the cumulative distribution plot presents a very different shape. 
Typically, smart Ponzi presents a small number of input transactions with few exceptions with many transactions.
The same happens for features \emph{Investment\_in/TX\_in}, \emph{Payment\_out/TX\_out} where smart Ponzi generally present larger values than the other class.
We expect that these features provide the classifier with a relevant contribution to discriminate between the two classes.
These differences are smoothed when we consider the features \emph{Payment\_out}, \emph{Percentage\_some\_tx\_in}, \emph{Percentage\_some\_tx\_out}, \emph{Sdev\_tx\_in}, and \emph{Sdev\_tx\_out} where the cumulative distributions are very similar. 
Therefore, we expect these features to contribute marginally to discriminating between the two classes.

\begin{figure}[t]
    \centering
    \includegraphics[width=\columnwidth]{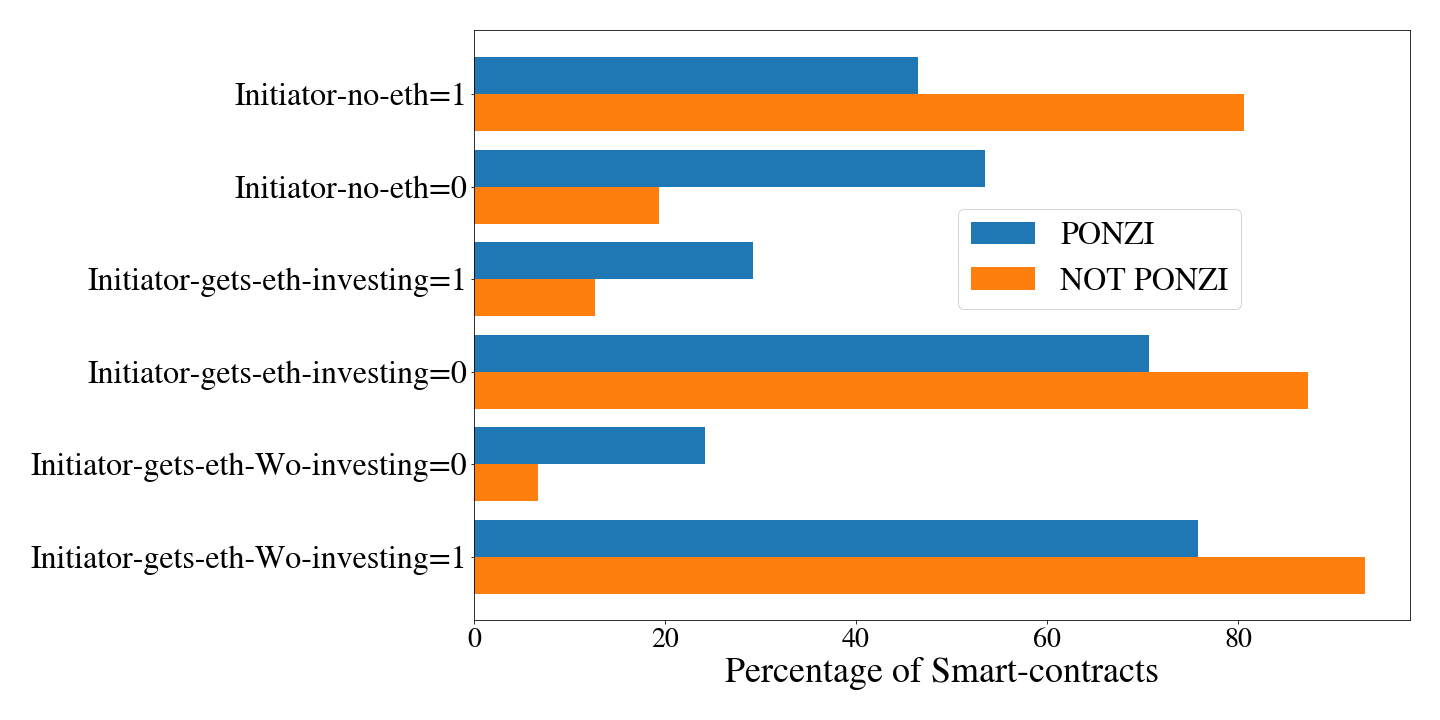}
    \vspace{-1.5em}
    \caption{The percentage of Ponzi and not Ponzi smart contracts for  Features 26-28.}
    \label{fig:binary}
\end{figure}

Figure~\ref{fig:binary} shows the distributions across the two classes in the percentage of the new binary features \emph{Initiator\_gets\_eth\_Wo\_investing}, \emph{Initiator\_gets\_eth\_investing}, \emph{Initiator\_no\_eth}. 
From the plot, we can see that the number of Ponzi contracts having this feature \emph{Initiator\_no\_eth} equal to 0 and 1 is similar. In contrast, most not-Ponzi contracts have this feature set to 1. 
We expect this can cause the classifier to make mistakes when discriminating between the two classes. 
Also, the feature \emph{Initiator\_gets\_eth\_investing} may not help the classifier since most contracts of the two classes have this feature equal to 0.
The feature \emph{Initiator\_gets\_eth\_Wo\_investing} could positively impact the classification because the number of not-Ponzi contracts with a value of 0 is low.

\section{Experiments}\label{sec:experiments}

In this section, we build binary classifiers for detecting smart Ponzi contracts. We perform an experimental evaluation to study how the new features of Section~\ref{sec:dataset} impact classification and the quality of the obtained classifiers. 
In particular, we answer the following three research questions:
\begin{description}
    \item[\textbf{RQ1:}] Do the new features 15-22 improve the classifier's quality? 

    \item[\textbf{RQ2:}] Which are the most relevant features? Can we find the best set of features?

    \item[\textbf{RQ3:}] Which are the characteristics of the best classifier?
\end{description}
Below, we report the experiments we performed to answer each question and their results. 
We carried out our experiments with Python, Scikit-learn, and the library SHAP~\cite{shapLib} for using XAI techniques.

\subsection*{RQ1: Impact of the new features}
We take the data sets \btwo\ and \paper\ and study the performances of classifiers trained with them. (Recall that \paper\ uses the same features of Chen et al.~\cite{Chen2019}.) 
In particular, we consider Decision Tree~\cite{quinlan1986induction}, Random Forest~\cite{breiman2001random}, and Light Gradient Boosting Machine Classifier (LGBMC)\cite{ke2017lightgbm} as classifiers and perform a grid search procedure with cross-validation to fine-tune the hyper-parameters of each classifier. 
We consider the Area Under the Curve (AUC) as the metric to be optimized.
We split the data sets into an 80\% training set (3537 samples) and a 20\% test set (885 samples) stratified on the target variable. 
Thus, given a training set, a classifier, and a combination of hyper-parameters, the cross-validation splits the training data into 5 folds. Then, the model (i.e., classifier and relative hyper-parameters) is trained and tested 5 times, varying the fold used as a validation set. 
An average of the metric to be optimized over the 5 tests is performed. The optimal classifier is the one with the highest mean score. In our case, the selected classifier has the highest mean AUC.
Once we have selected the best values for the hyper-parameters for each combination of data set and classifier, we compute the standard metrics \emph{Accuracy}, \emph{AUC}, \emph{F1}, \emph{Precision}, and \emph{Recall} on the test set.
Table~\ref{tab:comparison} reports a summary of the performances of the various classifiers on the two data sets.
According to the AUC metric, the best model for both data sets is LGBM but with different values for the hyper-parameters obtained through the grid-search procedure described above. For \btwo, the classifier has the following hyper-parameters: 80 estimators, max depth equal to 20, learning rate 0.1, 0.8 the subsample of columns, regulation alpha equal to 0.2 (L1 regularization term), and regulation lambda set to 1 (L2 regularization term).
For the data set \paper, the selected hyper-parameters are: 100 estimators, max depth equal to 15, learning rate 0.1, 0.5 the subsample of columns, regulation alpha equal to 0.1, and regulation lambda set to 10.
Moreover, the table shows that the model trained with \btwo\ presents a better AUC than the one trained with \paper. 
Therefore, we can answer \emph{positively} to RQ1:
\begin{mdframed}[style=answerbox, frametitle={Answer to RQ1:}]
\emph{The new features do improve the classifier's quality}. 
\end{mdframed}
Notice that for both data sets, we perform the grid search with the same values for the hyper-parameters, applying, thus, the same procedure to select the best-performing classifier.

\begin{figure}[t]
    \centering
    \includegraphics[width=\columnwidth]{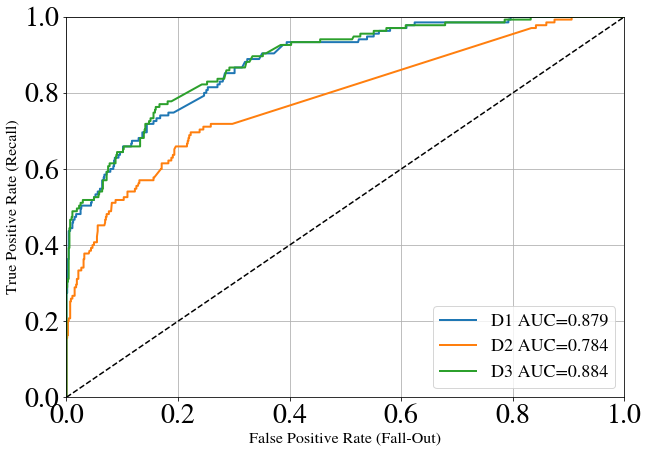}
    \caption{The ROC curve on the test set for the three best classifiers, one for each dataset. }
    \label{fig:roc_curve}
\end{figure}

Below, we analyze the results obtained by the LGBM classifier on \btwo\ and \paper.
Figure~\ref{fig:roc_curve} shows the ROC curve that describes the diagnostic ability of a classifier by comparing the true positive and the false positive rate. 
In Figure~\ref{fig:roc_curve}, the curve of \btwo\ presents higher values than the one of \paper, meaning that it exhibits a better classifier performance, as summarized in the fourth column of Table~\ref{tab:comparison}.

We also study the confusion matrices of the two models (see Figure~\ref{fig:confusion_matrices}).
%
From the matrices, we observe that the classifier trained with \btwo\ presents a higher number of TP and TN, i.e.~of smart contracts that are correctly classified, and a lower number of FP and FN, i.e.,~of smart contracts that are incorrectly classified.
Thus, \btwo\ performs better than \paper.


\begin{table}[tbp]
    \centering
    \scriptsize
    \caption{Results of the grid search comparing three classifiers on data sets \btwo\ and \paper (top). The performances on the best classifier on \bthree\ (bottom).}\label{tab:comparison}
\begin{tabular}{llccccc}
\toprule
      & Metric &  Accuracy &   AUC &    F1 &  Precision &  Recall \\
Data set & Classifier &           &       &       &            &         \\
\midrule
\btwo & Decision Tree &     0.855 & 0.733 & 0.496 &      0.529 &   0.467 \\
      & \textbf{LGBM} &     \textbf{0.904} & \textbf{0.879} & \textbf{0.608} &    \textbf{0.805} &   \textbf{0.489} \\
      & Random Forest &     0.896 & 0.875 & 0.562 &      0.787 &   0.437 \\
\paper & Decision Tree &     0.861 & 0.674 & 0.481 &      0.559 &   0.422 \\
      & \textbf{LGBM} &     \textbf{0.876} & \textbf{0.784} & \textbf{0.444} &   \textbf{0.698} &   \textbf{0.326} \\
      & Random Forest &     0.873 & 0.770 & 0.462 &      0.658 &   0.356 \\\midrule
\bthree      & \textbf{LGBM} &     \textbf{0.908} & \textbf{0.884} & \textbf{0.620} &   \textbf{0.846} &   \textbf{0.489} \\
\bottomrule
\end{tabular}
\end{table}

Finally, we consolidate our results through the McNemar test~\cite{mcnemar1947note}, a well-known approach to analyzing the statistical significance of the differences in classifier performances.
Our null hypothesis is that none of the two classifiers performs better than the other, whereas the alternative hypothesis is that the classifiers' performances are unequal.
We set the significance threshold to $0.05$ and compute the p-value, i.e., the probability of the observations where the null hypothesis is true.
The test results in a p-value equal to $0.0007$, thus lower than our significance threshold of $0.05$. 
Hence, the null hypothesis is rejected: the classifier trained on \btwo\ outperforms the one trained on \paper. This strengthens our answer to RQ1: \emph{the new features improve the classifier's quality, which is statistically significant and not due to some randomness in the data}. 

\begin{figure}[tbp]
\centering
\small
\begin{tabular}{l|cc|}
\multicolumn{3}{c}{Predicted class}\\
&N&P\\
\hline
N & $734$ & $16$\\
P & $69$ & $66$\\[1ex]
\multicolumn{3}{c}{Data set \btwo}
\end{tabular}
\qquad
\begin{tabular}{l|cc|}
\multicolumn{3}{c}{Predicted class}\\
&N&P\\
\hline
N & $731$ & $19$\\
P & $91$ & $44$\\[1ex]
\multicolumn{3}{c}{Data set \paper}
\end{tabular}
\qquad
\begin{tabular}{l|cc|}
\multicolumn{3}{c}{Predicted class}\\
&N&P\\
\hline
N & $738$ & $12$\\
P & $69$ & $66$\\[1ex]
\multicolumn{3}{c}{Data set \bthree}
\end{tabular}
\caption{Confusion matrices of the best classifier on our data sets, where we indicate with N and P the not Ponzi and Ponzi class respectively.}
\label{fig:confusion_matrices}
\end{figure}

\begin{figure}[tbp]
	\centering
    \includegraphics[width=\columnwidth]{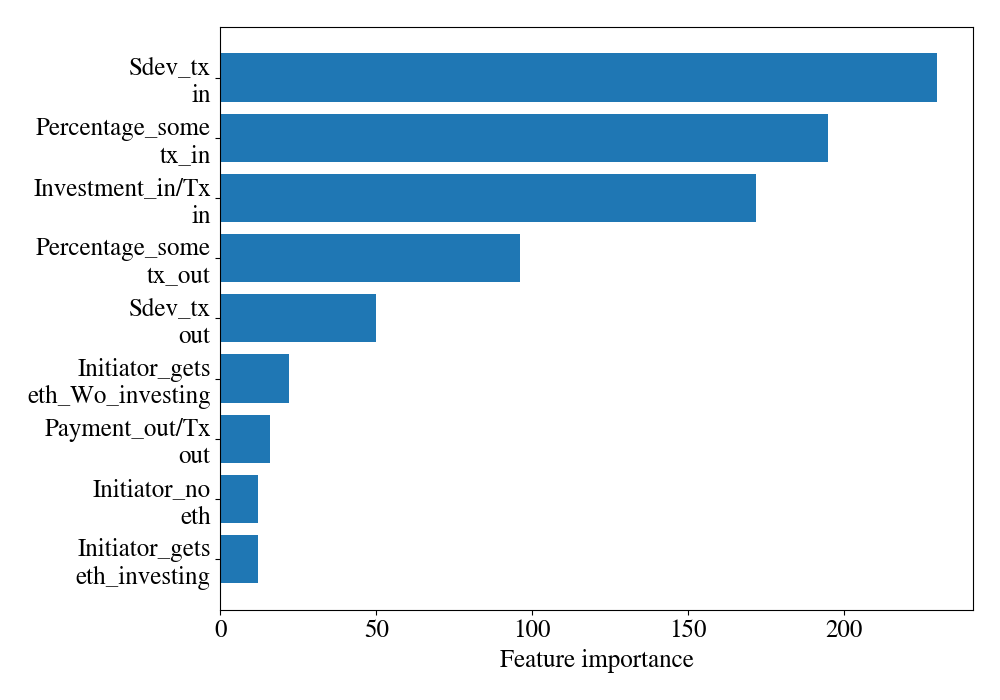}
    \vspace{-4mm}
    \caption{The importance of the new set of features included in dataset \btwo.}
    \label{fig:featureimportance}
    \vspace{-5mm}
\end{figure}

\subsection*{RQ2: Detect most relevant features}
We take our best model on \btwo\ and investigate which features contribute most to classification and which mislead the classifiers.
First, we compare the new features to determine their importance. 
Each classifier adopts different metrics to determine the importance of features.
The LGBM classifier considers the number of times a feature is used to split the data across all ensemble trees: the more a feature is used, the more important it is.
Figure~\ref{fig:featureimportance} shows the new features sorted by importance, where the x-axis represents the number of splits. 
We see that \emph{Sdev\_tx\_in} is the most important, whereas \emph{Initiator\_no\_eth}
and \emph{Initiator\_gets\_eth\_investing} are less important. 

Then, we consider the problem of determining if there exists a subset of the features of \btwo\ that improves the quality of the classifier. 
Our idea is to consider the number of features as a further hyper-parameter to be optimized. 
To do that, we proceed as follows.
We perform a grid search procedure with cross-validation to optimise the AUC and the number of features. 
To tune this last hyper-parameter, we start considering all the features in \btwo, perform a grid search and 5-fold cross-validation procedure with the other hyper-parameters and obtain the best-performing classifier. Then, we adopt the \emph{Recursive Feature Elimination} algorithm to remove the less important feature. Given the new reduced data set, the procedure is repeated until the number of features to be evaluated equals the number of features of \paper. 
As for the previous experiments, we split the data set into an 80\% training set (3537 samples) and a 20\% test set (885 samples).
As a result, we obtain the highest mean AUC with the data set \bthree\, which includes 25 features (see Table~\ref{tbl:dataset:features}). 
The LGBM classifier trained on \bthree\ performs better and improves the best classifier on \btwo\ as shown in Table~\ref{tab:comparison}. 
The other hyper-parameters are the following:  120 estimators, max depth equal to 15, learning rate 0.1, 0.5 the subsample of columns, regulation alpha equal to 0.1, and regulation lambda set to 10.
In \bthree\ the following features are removed in order \emph{Initiator\_no\_eth}, \emph{Initiator\_gets\_eth\_investing}, and \emph{Payment\_out}. This confirms what was anticipated in Section~\ref{sec:dataset}:
\emph{Initiator\_no\_eth} and \emph{Initiator\_gets\_eth\_investing} do not contribute to discriminating between the two classes because creators of legit smart contracts may receive or not an amount of money from contracts they created.
This is in line with requirements R1-R4 of Section~\ref{sec:back} that do not consider the money received by the initiator as a discriminating factor to be a Smart Ponzi.
Finally, \emph{Payment\_out} does not contribute to classification because every smart contract transfers money during its lifecycle. 
Thus, the number of output transactions on its own seems not to capture requirement R1 of Section~\ref{sec:back}, so it is not a good estimator to detect Ponzi schemes.
In summary:
\begin{mdframed}[style=answerbox,frametitle={Answer to RQ2:}]
\emph{The best and relevant features are those used by dataset \bthree.}
\end{mdframed}

\subsection*{RQ3: Characteristics of the best classifier}

\begin{figure}[tbp]
    \centering
    \includegraphics[width=\columnwidth]{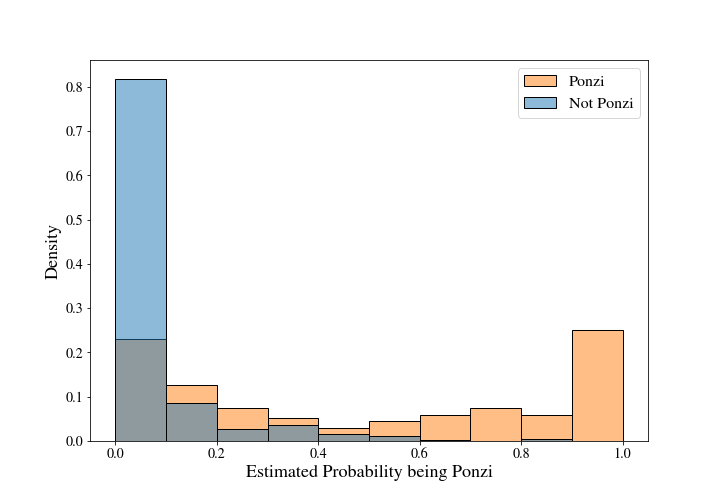}
    \vspace{-4mm}
    \caption{Distribution of the probability to be a Ponzi scheme estimated using the best-performing classifier on \bthree. We report the distribution of both classes.}
    \label{fig:probability}
    \vspace{-2mm}
\end{figure}

\begin{figure}[t]
    \centering
    \includegraphics[width=\columnwidth]{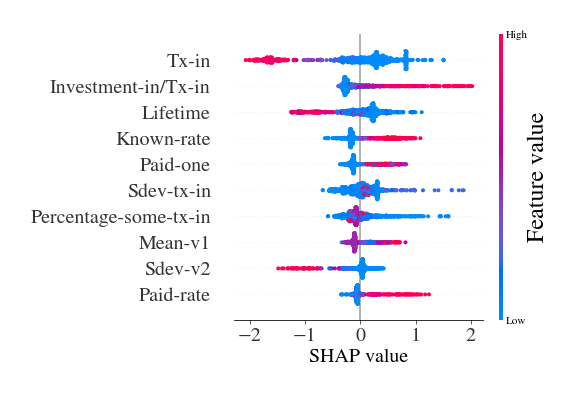}
    \vspace{-4mm}

    \caption{The feature importance of the top 10 features in terms of SHAP values.}
    \label{fig:summary_shap}
\end{figure}

As described above, the best classifier is LGBM, considering the 25 features of \bthree. 
Figure~\ref{fig:confusion_matrices} (right) shows the corresponding confusion matrix: 4 false positive contracts are now classified correctly. 
Figure~\ref{fig:probability} displays histograms of the probability of being in one of the two classes. 
Considering a threshold of $0.5$, the classifier may assign a lower probability to some smart Ponzi contracts. 
To understand how the classifier works on \bthree, we use the XAI library SHAP to investigate how the most important features impact the classification. 
Figure~\ref{fig:summary_shap} shows the beeswarn plot of the feature importance based on SHAP values for the test set. 
The plot displays an information-dense summary of how the top features impact the model output. 
Each instance is represented by a single dot whose position on the $x$-axis is determined by the SHAP value of the feature; the dots are ``pile up'' along each feature row to show the density. 
Colour is used to display the original value of a feature: blue corresponds to a lower value, while red corresponds to a higher value. 
Below, we comment on the top three features.
We can easily understand each feature's positive or negative effect on the prediction outcome from the plot. 
We observe that the number of transactions in input (feature Tx\_in) negatively impacts when it presents a high value. 
This means a small number of input transactions probably characterizes smart Ponzi contracts. 
We cannot explain why this happens only by observing this feature in isolation but by considering its relation with others (see the discussion below). 
On the contrary, a higher value of feature
\textit{Investment\_in/tx\_in} indicates a positive effect on the classifier to label an instance as a smart Ponzi. 
This could be because typical smart Ponzi contracts do not provide users with many services besides investing some money. 
Thus, most input transactions the contract receives carry a certain amount of currency. 
The \textit{Lifetime} feature helps to discriminate between the two classes because the time smart Ponzi contracts are active is typically short.
Indeed, they may quickly reach the point where it becomes difficult for users to have revenue, so they collapse. 
This behaviour is evident from the plot of Figure~\ref{fig:summary_shap} where contracts with smaller values of the \textit{Lifetime} feature tend to be classified as smart Ponzi ones.
The importance of the rest of the features could be explained similarly.
A last comment on the results of Figure~\ref{fig:summary_shap} is that three of the features introduced in Section~\ref{sec:dataset} belong to the top 10.  
This confirms our choice of adding them to improve the classification. 

\begin{figure*}[p]
    \centering
    \footnotesize
    \begin{tabular}{cccc}
    \includegraphics[width=0.85\columnwidth]{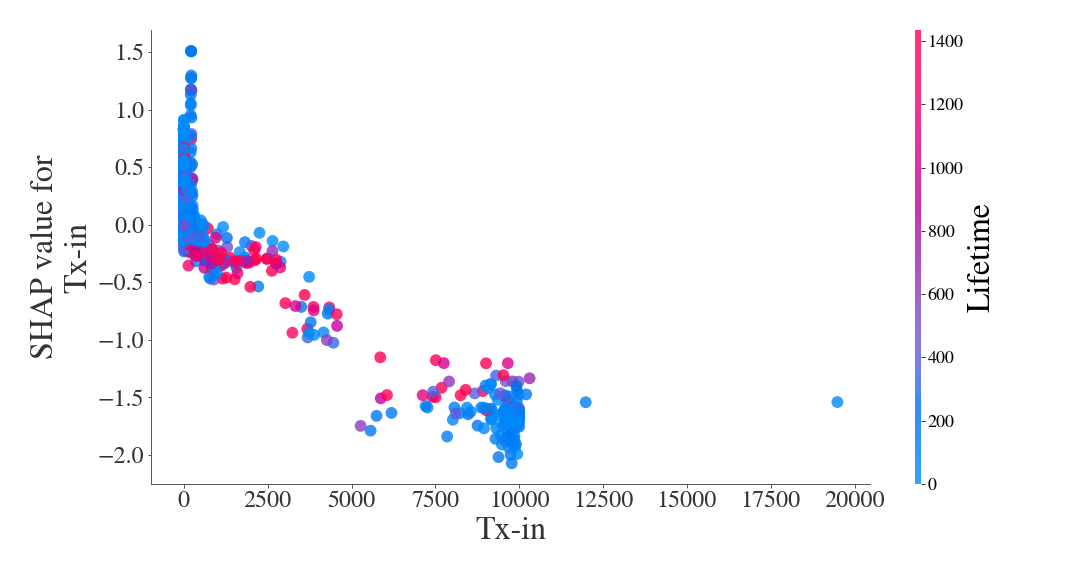} & 
    \includegraphics[width=0.85\columnwidth]{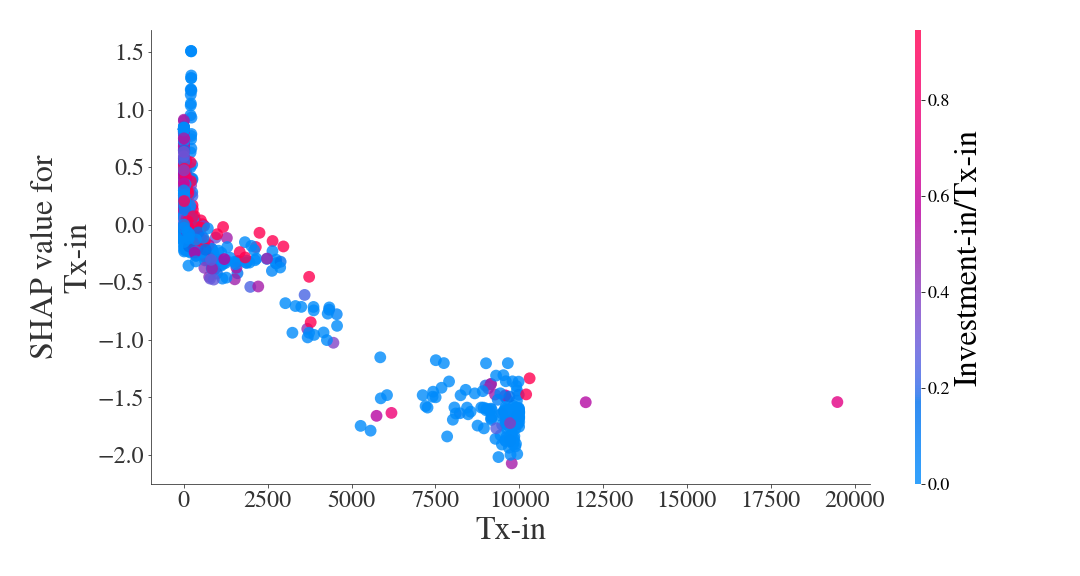} \\
    \includegraphics[width=0.85\columnwidth]{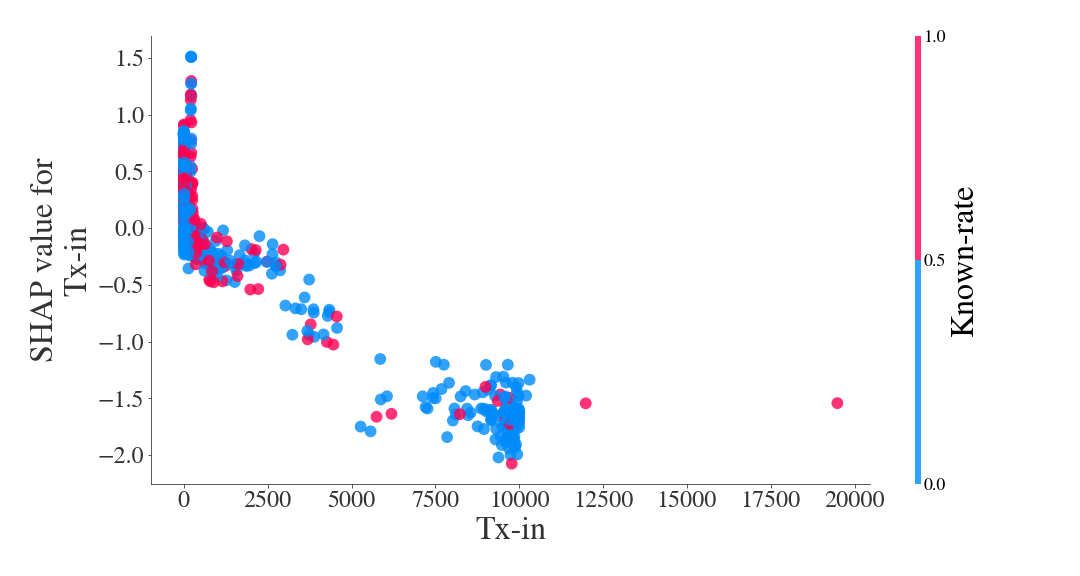} &
    \includegraphics[width=0.85\columnwidth]{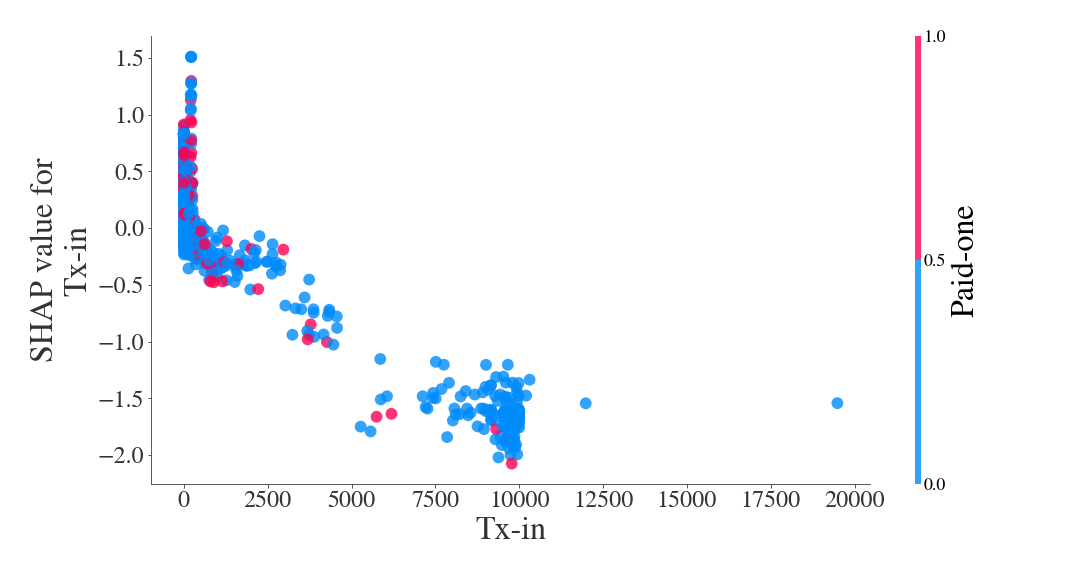} \\
    \includegraphics[width=0.85\columnwidth]{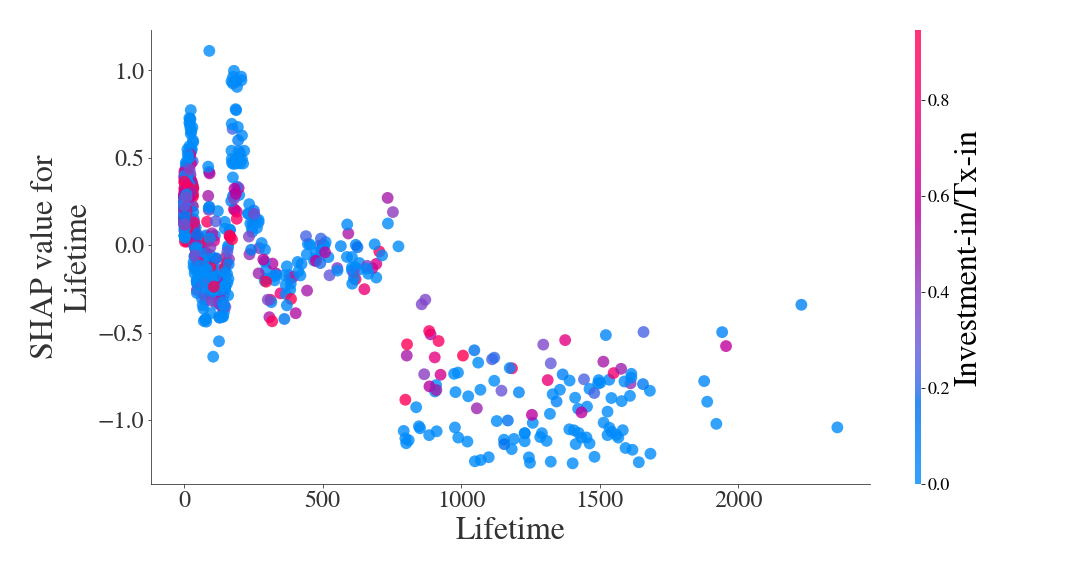} &
    \includegraphics[width=0.85\columnwidth]{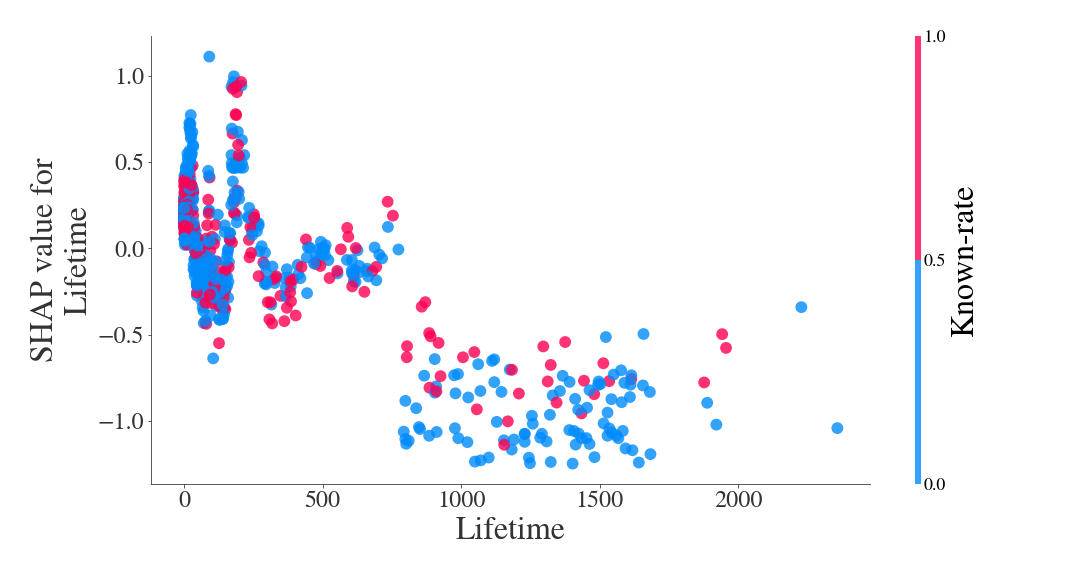} \\
    \includegraphics[width=0.85\columnwidth]{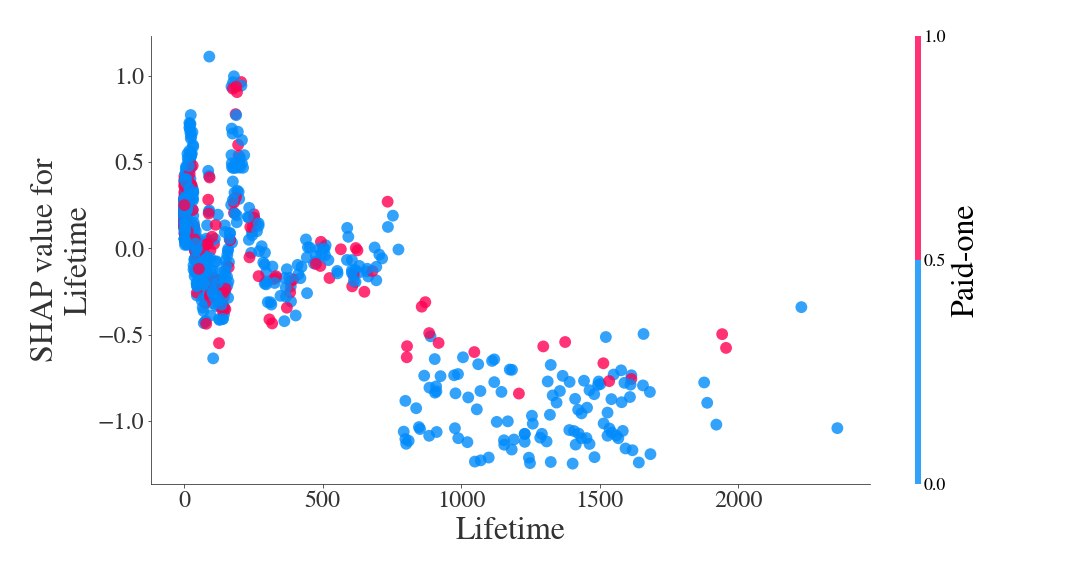} & 
    \includegraphics[width=0.85\columnwidth]{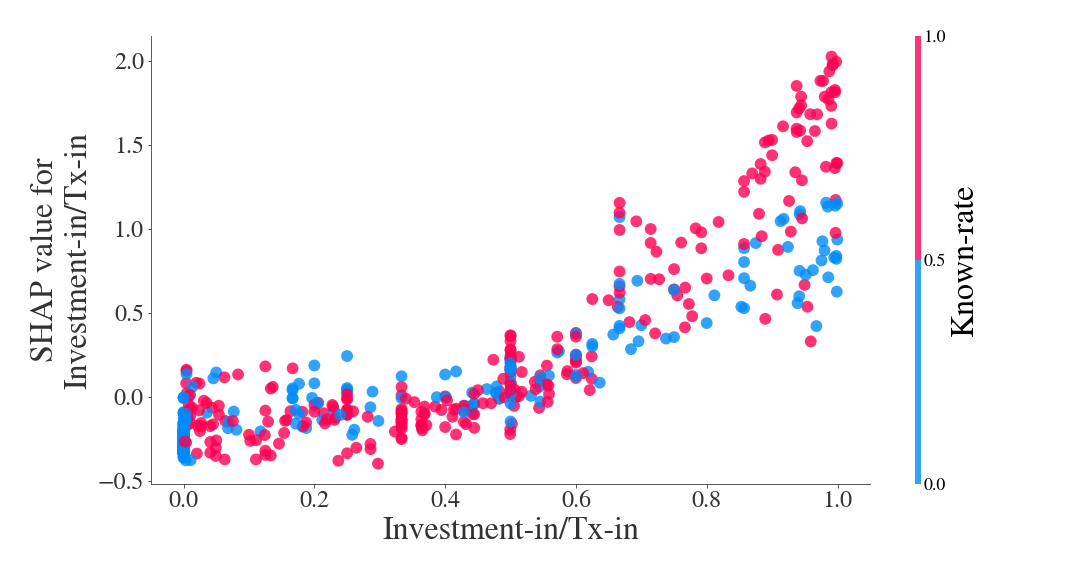} \\
    \includegraphics[width=0.85\columnwidth]{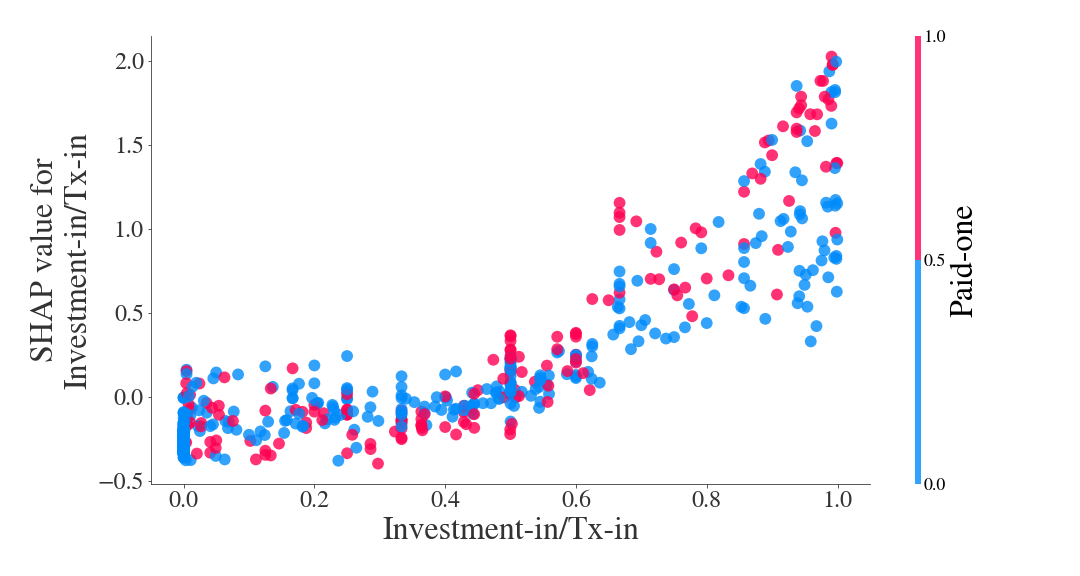} & 
    \includegraphics[width=0.85\columnwidth]{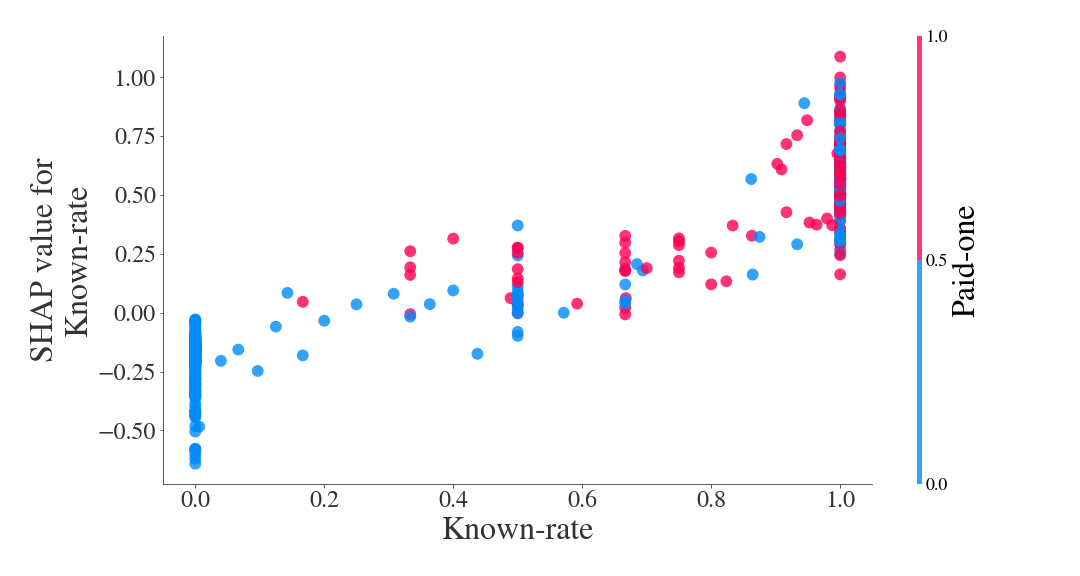} \\
    \end{tabular}
    \caption{The dependence plots for the five most important features: Tx\_in, Investiment-in/Tx-in, Lifetime, Known rate and Paid one.}
\label{fig:dependency:plots}
\end{figure*}
As we already anticipated, a single feature in isolation cannot capture requirements R1-R4, but a group of features is required. 
So, the interaction among features plays an important role in classifying a contract as a smart Ponzi or not. 
To study and explain these interactions, we report in Figure~\ref{fig:dependency:plots} the dependence plots generated through the SHAP library for the five most important features of Figure~\ref{fig:summary_shap}: Tx\_in, Investiment-in/Tx-in, Lifetime, Known-rate and Paid one. 
Intuitively, a dependence plot is a scatter plot that shows the effect of a single feature on the predictions made by the model.  
In the plot, each dot represents a sample (a smart contract in our case), the $x$-axis represents the feature values, the $y$-axis is the SHAP value for that sample, and the colour intensity indicates the values of a second feature on which we want to study the dependency.  
A distinct vertical pattern of colouring suggests a certain level of interaction or dependency between the considered features.
In our case, the second feature is always taken from the most important ones.
The first plot on the left of Figure~\ref{fig:dependency:plots} highlights a certain interaction between Tx\_in and Lifetime to predict the probability of being a Ponzi scheme.
More precisely, we observe that when the value of Tx\_in is small, the contract's lifetime can be small too, and the corresponding SHAP value can span in a wide range, showing the vertical colouring pattern that indicates the interaction between features.
This dependency between the number of input transactions and the lifetime is clear when considering what we already anticipated above. 
Indeed, smart Ponzi contracts are typically characterized by a small number of input transactions and by a short lifetime because it is probable that they collapse quite soon. 
A similar pattern emerges when we consider Lifetime with the other features such as Investment-in/Tx-in, Know-rate and Paid-one. 
When the Lifetime value is low, the values of the second feature present a certain density of the same colour, whereas when the value of the Lifetime is low, the other features present different colours. 
Moreover, from the plots, we observe the interaction between Lifetime and Investment-in/Tx-in ratio is similar to the one between Lifetime and Paid-one. 
This may be because the number of transactions is a function of the number of investors. 
The plots show another clear interaction between the features Investiment-in/Tx-in and Know-rate and Paid-one. 
In particular, when the value of the Investiment-in/Tx-in ratio is high, a vertical colour pattern emerges for high/low values of Know-rate/Paid-one. 
The interactions are less evident for the other combinations of features, and we do not show them.   

In summary:
\begin{mdframed}[style=answerbox,frametitle={Answer to RQ3:}]
\emph{The analysis of the best classifier reveals that smart Ponzi contracts are characterized by a short lifetime, by a small number of input transactions that provide a high income of money, and by a number of transactions that pay a small subset of investors.
Therefore, our results are in line with the requirements R1-R4.}
\end{mdframed}

\section{Conclusion}\label{sec:concl}

This paper presented an automatic technique for detecting smart Ponzi contracts on Ethereum.
We released a reusable data set with 4422 unique real-world smart contracts that can be used for future research.
Then, we introduced a new set of features that allowed us to improve the classification and outperform previous efforts in the literature~\cite{Chen2019}.
Finally, we identified a small and effective set of features that ensures a good classification quality, and we applied XAI techniques to explain how the most important features impact classification.

In future work, we plan to extend our model in different directions. We intend to improve the procedure to optimize the best set of features, we would like to take into account also the bytecode of contracts present in our dataset but not used by our current classifier and possibly apply deep learning techniques to minimize the feature engineering effort.
Moreover, we plan to further investigate the relationship among the most important features and to refine the requirements of being a smart Ponzi proposed by Bartoletti et al. as well the temporal dimension of such a set of features. 
Analogously, our approach can be applied to detect other forms of scams on Ethereum, and phishing is one of the most promising. 

\section*{Acknowledgment}

This work was partially supported by project SERICS (PE00000014)
under the MUR National Recovery and Resilience Plan funded
by the European Union - NextGenerationEU
\bibliographystyle{ACM-Reference-Format}
\bibliography{biblio}

\end{document}